# Topological Magnetoelectric Response in Ferromagnetic Axion Insulators

Yuhao Wan[1], Jiayu Li[1], and Qihang Liu[1,2,3,*]

[1]*Department of Physics and Shenzhen Institute for Quantum Science and Engineering (SIQSE), Southern University of Science and Technology, Shenzhen 518055, China*

[2]*Shenzhen Key Laboratory of Advanced Quantum Functional Materials and Devices, Southern University of Science and Technology, Shenzhen 518055, China*

[3]*Guangdong Provincial Key Laboratory for Computational Science and Material Design, Southern University of Science and Technology, Shenzhen 518055, China*

Y.W. and J.L. contributed to the work equally.

[*]Email: liuqh@sustech.edu.cn

## Abstract

Topological magnetoelectric effect (TME) is a hallmark response of the topological field theory, which provides a paradigm shift in the study of emergent topological phenomena. However, its direct observation is yet to be realized due to the demanding magnetic configuration required to gap all surface states. Here, we theoretically propose that axion insulators with a simple ferromagnetic configuration, such as $MnBi_2Te_4/(Bi_2Te_3)_n$ family, provide an ideal playground to realize TME. In the designed triangular prism geometry, all the surface states are magnetically gapped. Under a vertical electric field, the surface Hall currents give rise to a nearly half-quantized orbital moment, accompanied by a gapless chiral hinge mode circulating in parallel. Thus, the orbital magnetization from the two topological origins can be easily distinguished by reversing the electric field. Our work paves a new avenue towards the direct observation of the TME in realistic axion-insulator materials.

**Keywords:** topological magnetoelectric effect, axion insulator, magnetic topological materials

**Introduction**

Topological magnetoelectric effect (TME), *e.g.*, the topological response of magnetization to an electric field in the same direction, is a hallmark phenomenon of the topological field theory [1, 2]. Topological materials that possess a three-dimensional (3D) bulk axion field $\theta = \pi \pmod{2\pi}$ accompanied by surface energy gaps are designated as axion insulators (AXIs) [1, 3-7], which are expected to exhibit TME. However, observing the TME is highly challenging owing to the requirement of introducing magnetic gaps on all the surfaces of a system with bulk $\theta = \pi$ [8-10]. The recent advent of intrinsic magnetic topological insulators (TIs), especially the MnBi$_2$Te$_4$/(Bi$_2$Te$_3$)$_n$ family [11-25], has provided an ideal platform for the AXI phase. To date, the verification of the axion state has mostly focused on the measurement of the combined effect of two surfaces, say the top and the bottom, such as the quantized Faraday/Kerr rotations [26-28] and the zero Hall plateau (ZHP) [9, 29-32]. While the ZHP is conventionally regarded as a hallmark of AXIs, it is not exclusive to these, and ZHP signals may also appear in topological thin films gapped by quantum confinement and other non-axion cases [25, 33-37]. Recent progress has focused on signals related to the surface anomalous Hall effect at a single gapped surface [37-43], which directly reveals the bulk-boundary correspondence of AXIs. Nevertheless, TME, as the central assertion of topological field theory, is yet to be observed.

The difficulty in realizing TME mainly lies in the requirements of geometric and magnetic configurations, which are quite challenging for realistic materials. Historically, it was proposed that TME could be realized in a spherical strong TI with a hedgehog-like magnetization on its surface, or a cylindrical TI with a surface radial magnetization, as shown in Fig. 1(a) and (b), respectively [1, 4, 7]. Unfortunately, neither of these over-idealized configurations are accessible for experiments. It is conventionally believed that TME can only be realized in a configuration without a net magnetic moment, *e.g.*, antiferromagnetic (AFM) systems [4-6]. The reason is that a ferromagnetic (FM) AXI is typically a higher-order topological insulator (HOTI) simultaneously, accompanied by 1D hinge modes or magnetization induced quantum

Hall effects [see Fig. 1(c)] [1, 44-47]. However, for AFM AXIs such as MnBi$_2$Te$_4$, the side surface state is gapless, protected by the combined symmetry of time reversal $\mathcal{T}$ and fractional translation $\tau_{1/2}$ [14, 15], which also obscures the observation of TME.

Here, we theoretically demonstrate that by designing particular configurations, an FM AXI, such as MnBi$_2$Te$_4$/(Bi$_2$Te$_3$)$_n$, can serve as an ideal platform for realizing a topological magnetoelectric response. The simple magnetic configuration, which could be achieved directly in MnBi$_8$Te$_{13}$ with an FM ground state [21, 40] and in MnBi$_2$Te$_4$, MnBi$_4$Te$_7$, and MnBi$_6$Te$_{10}$ (AFM ground state) [13, 16, 20, 22] under a moderate vertical magnetic field, ensures the accessibility of our proposal. Remarkably, the side surface states of such FM AXIs manifest magnetic gaps owing to the hexagonal warping effect [19, 48]. In the designed geometry of the triangular prism [see Fig. 1(d)], a chiral hinge mode from the HOTI phase is pinned to circulate around the top surface rather than the entire bulk. When an electric field is applied along the prism, the side surface Hall current also circulates parallel to the top surface, thus avoiding interference with the hinge mode. We calculate the bulk magnetization as a response to the external electric field and obtain the nearly half-quantized response coefficient. In a realistic finite-layer system, the TME-induced magnetization can be directly extracted by reversing the electric field, as the signals from the TME and hinge mode are odd and even under field reversal, respectively. Moreover, we characterize the thickness-driven crossover of the FM AXI from 3D HOTI to 2D Chern insulator by visualizing the distribution of the chiral mode. Our findings provide an accessible material-based proposal to achieve the long-sought TME to validate the topological field theory.

**Configuration to realize topological magnetoelectric effect**

According to the topological field theory, the gradient of the static axion field $\nabla\theta$ at the surfaces of TIs leads to TME through the surface half-quantized Hall current [1, 3]. Imagining a 3D time-reversal invariant TI with surface states gapped homogeneously by spatially oriented surface magnetic moments [see Figs. 1(a) and 1(b)], the surface Hall current density induced by the external electric field **E** is

$$\mathbf{j}_H = \frac{\alpha}{\pi} \mathbf{E} \times \nabla \theta, \quad (1)$$

with $\alpha = e^2/2hc$ the fine structure constant. Such a circulating current contributes to TME, *i.e.*, an emergent magnetization parallel to **E** with a quantized coefficient $\mathbf{M}_{TME} = \alpha\mathbf{E}$. Reexamining the necessity of such impractical spatially-oriented magnetization, we find that under FM moments, the shell of the spherical TI is divided into two domains with opposite magnetic surface gaps forming a gapless chiral mode at the domain wall. In most cases, dissipation occurs when the circulating currents encounter the gapless chiral mode, violating the adiabatic condition of TME, *i.e.*, the side surface states should be gapped to ensure the circulation of the Hall currents [1, 9]. However, if the chiral mode is pinned to circulate around a 2D plane, interference between TME and the chiral mode would be avoided if the electric field is applied perpendicular to the plane. Although in Fig. 1(c) the surface Hall currents from the two hemispheres compensate each other, leading to a zero net magnetoelectric response, it inspires us that FM systems could potentially reveal the TME by designing inequivalent domains.

Exemplified by $MnBi_2Te_4/(Bi_2Te_3)_n$, we propose a material-based structure to realize TME in an FM AXI with triangular prism geometry, as shown in Fig. 1(d). Previous studies have shown that the $MnBi_2Te_4$ family of out-of-plane FM ordering are both AXIs protected by inversion symmetry and HOTIs [19, 21, 41]. In a hexagonal prism sample, the six side surfaces are gapped by staggered mass terms with respect to the three-fold rotation and inversion symmetries [19, 46]. The gapless chiral modes at the domain walls of the side surfaces thus obscure the measurements of TME. However, once the AXI adopts a triangular prism configuration, all three side surfaces have the same sign of magnetic surface gaps, which is coincident with that of, say, the bottom surface. Therefore, the gapless chiral mode is localized only at the top surface, rather than circulating around the entire bulk. Once the vertical electric field is applied, the responding Hall current travels parallel to the hinge mode without any interference, leading to a measurable magnetization (anti)parallel to the electric field. Recently, the HOTI bismuth with similar helical hinge modes [49] was successfully synthetized in a triangular geometry [50]. This implies that our proposal for realizing TME in the FM MnBi$_2$Te$_4$ family is experimentally accessible.

**Model for ferromagnetic axion insulator MnBi$_2$Te$_4$**

To calculate the TME of FM AXI, we start from the effective model Hamiltonian of FM MnBi$_2$Te$_4$ written in a triangular lattice [19, 42]

$$\mathcal{H} = d_0 I_4 + \sum_{i=1,\cdots,5} d_i \Gamma_i + \Delta \mathbf{m} \cdot \mathbf{s} \otimes \sigma_0, \tag{2}$$

where $d_0 = \tilde{C} - 2C_1 \cos k_z - (4C_2/3)(\cos k_1 + \cos k_2 + \cos k_3)$, $d_1 = (v/3)(2\sin k_1 + \sin k_2 + \sin k_3)$, $d_2 = (v/\sqrt{3})(\sin k_2 - \sin k_3)$, $d_3 = v_z \sin k_z$, $d_4 = 8w(-\sin k_1 + \sin k_2 + \sin k_3)$, and $d_5 = \tilde{M} - 2M_1 \cos k_z - (4/3)M_2(\cos k_1 + \cos k_2 + \cos k_3)$ with $\tilde{R} = R_0 + 2R_1 + 4R_2$ ($R = C, M$), $k_1 = k_x$, $k_2 = (k_x + \sqrt{3}k_y)/2$, and $k_3 = k_1 - k_2$. Here $I_4$ is the identity matrix, $\Gamma_i = s_i \otimes \sigma_1$ for $i = 1,2,3$, $\Gamma_4 = s_0 \otimes \sigma_2$, and $\Gamma_5 = s_0 \otimes \sigma_3$, where $s_i$ and $\sigma_i$ are the Pauli matrices for spin and orbital, respectively. $v$ ($v_z$) represents the velocity along the in-plane (out-of-plane) direction, $w$ is the hexagonal warping parameter [48, 51], $\Delta$ describes the exchange coupling between the electron states and magnetic moments, and $\mathbf{m} = (0,0,1)$ for the out-of-plane FM order. The model parameters are presented in Supplementary data Section 1.

The model Hamiltonian of FM MnBi$_2$Te$_4$ respects three-fold rotation around the $z$-axis $C_{3z} = e^{i\pi s_3/3} \otimes \sigma_0$ and inversion $\mathcal{P} = s_0 \otimes \sigma_3 = \Gamma_5$. Consequently, the bulk topology of this inversion-preserved system can be described by the symmetry indicator of $\mathcal{P}$, i.e., $\mathbb{Z}_4 \times \mathbb{Z}_2 \times \mathbb{Z}_2 \times \mathbb{Z}_2$ [52-55]. We thus verify that the FM MnBi$_2$Te$_4$ is an FM AXI with a symmetry indicator (2;000) (Supplementary data Section 1). Such an AXI phase can also be confirmed by calculating the real-space resolved Chern marker [56-58]. In Fig. 2(a) we show that in the 16-slab model, the top and bottom four layers contribute a nearly half-quantized Chern number $C_{t(b)} = \pm 1/2$, indicating the bulk topology $\theta = \pi$. In addition, owing to the hexagonal warping term $d_4$, the side surface states open a magnetic gap (Supplementary data Section 1). Previous first-principles calculations show that in FM MnBi$_2$Te$_4$ such a high-order magnetic gap is about 6 meV [41], which is typically larger than the finite-size gap of the few-layer slabs. Since the top and bottom surface states are gapped by the exchange coupling, all the surfaces of the configuration are gapped. Due to the $C_{3z}$ symmetry, the gap signs of the side surfaces are the same and coincides with either that of the top surface or the bottom surface (Supplementary data Section 1). In this work, $(100), (0\bar{1}0), (\bar{1}10)$ surfaces are chosen as the side surfaces, of which the bottom surface shares the same gap sign.

**Topological magnetoelectric response**

To evaluate the TME, we next calculate the orbital magnetization induced by the vertical electric field through linear response theory. By constructing an equilateral triangle lattice in the $xy$-plane with a side length of $L$ sites while maintaining the translation symmetry along $z$, we obtain the field-induced orbital magnetization $M_{TME,i} = \alpha_{ij}E_j$, in which the only relevant coefficient component $\alpha_{zz}$ is calculated via the Kubo formula [59]

$$\alpha_{zz} = \frac{-i}{V}\lim_{\Omega \to 0}\int_{-\infty}^{+\infty}dt' e^{i\Omega(t-t')}\theta(t-t')\langle[\widehat{M}_z(t),\hat{J}_z(t')]\rangle, \qquad (3)$$

where $V$ is the system volume, $\theta(x)$ is the step function, and $\Omega$ is the frequency of the electric field. $\langle O \rangle = \text{Tr}[e^{-\beta H}O]/\text{Tr}e^{-\beta H}$ denotes the ensemble average with $\beta$ the inverse temperature. The orbital magnetic moment is defined as $\widehat{\mathbf{M}} = -(e/2c)\hat{\mathbf{r}}\times\hat{\mathbf{v}}$ [60], with $\hat{\mathbf{r}}$ and $\hat{\mathbf{v}}$ the position and velocity operator, respectively, and the current density is $\hat{\mathbf{J}} = e\hat{\mathbf{v}}$. In the zero-temperature limit with the Fermi level $\mu_F$ lying inside the band gap, Eq. (3) can be simplified as follows [61]

$$\alpha_{zz} = \frac{2\hbar}{V}\sum_{k_z}\sum_{\substack{i\in occ.\\ j\in unocc.}}\frac{Im[\langle k_z,i|\widehat{M}_z|k_z,j\rangle\langle k_z,j|\hat{J}_z|k_z,i\rangle]}{(\varepsilon_{i,k_z}-\varepsilon_{j,k_z})^2}, \qquad (4)$$

where $|k_z,i\rangle$ is the $i$-th eigenstate with momentum $k_z$ with respect to the eigenvalue $\varepsilon_{i,k_z}$. During the summation, $i$ and $j$ are constrained within the occupied and unoccupied bands, respectively.

The numerical results of the magnetoelectric response coefficient $\alpha_{zz}$ for different side lengths $L$ are shown in Fig. 2(b). If the system size is large enough ($L \geq 12$), $\alpha_{zz}$ approaches a nearly quantized value $0.9\alpha$ with an exchange coupling $\Delta = 62$ meV. This is the central result of our work. To verify our theoretical approach and the topological origin of $\alpha_{zz}$, we tune the parameter $M_0$ in $d_5$ to drive the MnBi$_2$Te$_4$ bulk to the $\theta = 0$ side, and find that the corresponding $\alpha_{zz}$ vanishes accordingly (Supplementary data Section 2). According to the $\theta = \pi$ nature in a bulk AXI, we can expect a quantized $\alpha_{zz} = \alpha$ in a large size AXI sample with all the surface gapped and preserved symmetry (e.g., inversion symmetry) that protects the bulk axion field. This can be verified by calculating the TME of an ideal 3D AXI protected by both $\mathcal{P}$ and $\mathcal{T}$ with a cubic lattice and radial side surface magnetization (Supplementary data Section 2). Even $\mathcal{T}$ symmetry is broken by the surface magnetization, the finite tetragonal structure preserves $\mathcal{P}$ symmetry and we find that $\alpha_{zz}$ approaches to $\alpha$ for increasing. Therefore, in our FM AXI configuration, the deviation between the saturation value $0.9\alpha$ and the perfect quantization ($\alpha$) is attributed to the inversion breaking of the triangular prism.

The profile of $\alpha_{zz}$ experiences a steep increase followed by oscillation and eventually saturation as the magnetization $\Delta$ increases from zero. To understand this, we plot the band spectra along $k_z$ for different $\Delta$ values, as shown in the inset of Fig. 2(b). Without magnetization ($\Delta = 0$), the band gap, *i.e.*, the side surface gap of the triangular prism, originates from the finite size effect of the in-plane triangular geometry characterized by $L$, whereas for $\Delta = 70$ meV the gap is dominated by magnetization after a nontrivial phase transition. Here the nontrivial topology of this finite size system is characterized by the topological magnetoelectric coefficient, *i.e.*, nonzero Hall current induced by the external electric field. Although there is no gap closing accompanied as $\Delta$ increases, the phase transition could be monitored by the change in the local Chern marker of the side surface (Supplementary data Section 3), thus giving rise to the evolution of $\alpha_{zz}$. As $L$ increases, $\alpha_{zz}$ grows more rapidly when turning on $\Delta$. It is because the finite-size induced hybridization gap reduces and thus becomes easier to be overwhelmed by the magnetic gap.

In realistic samples, the translation symmetry along the $z$ direction is broken and the prism is terminated by the top and bottom surfaces, which have a larger magnetic gap than those of the side surfaces. Consequently, the top surface carries a gapless chiral hinge mode [see Fig. 1(d)], while there is no hinge mode propagating along the $z$ direction. We consider several prism models with different heights of $H$ layers and directly diagonalize them, and find that the energy of the hinge state converges as $H > 10$. The distribution of the corresponding wavefunction decays exponentially from the hinge (Supplementary data Section 4). Figs. 3(a-c) show the prism geometry, energy level, and distribution of the hinge state of the FM AXI with $L = 10$ and $H = 16$. Such a hinge state does contribute an extra orbital magnetization $M_{IC} \simeq -e\mu_F/\hbar c$, which is also known as itinerant circulation magnetization [60, 62].

Next, we discuss how to distinguish the TME-induced orbital moment from the total magnetization. Since there is no interference between the surface Hall current and the chiral hinge mode, the total bulk magnetization $M_{tot}$ is a superposition of the local moment $M_{ion} = n_i \langle S_{Mn} \rangle$ from Mn ions ($n_i$ and $\langle S_{Mn} \rangle$ are the density and average spin of Mn ions, respectively), itinerant circulation magnetization $M_{IC} \simeq -e\mu_F/\hbar c$ from the chiral hinge state, and the TME-induced magnetization $M_{TME} = \alpha_{zz} E_z$ from the surface Hall current. Both $M_{ion}$ and $M_{IC}$ are independent on the vertical electric field (up to the first order, before it leads to a topological phase transition), whereas $M_{TME}$ is an odd function of **E**. Therefore, the net TME signal can be easily extracted via two measurements with opposite electric fields $M_{TME} = [M_{tot}(\mathbf{E}) - M_{tot}(-\mathbf{E})]/2$, as illustrated in Fig. 3(d).

**Magnetic gap versus hybridization gap**

For a thin slab (small $H$), the hinge mode encounters the bottom surface before it completely evanesces, leading to a crossover between 3D HOTI and 2D Chern insulator, as shown in Fig. 4. Both the HOTI and Chern insulator phases manifest half-quantized anomalous Hall conductivities at the top and bottom surfaces [4, 41]. The crossover between such 3D and 2D topological matter is determined by the gap nature of the side surface, *i.e.*, the competition between magnetization and finite-size hybridization. With large enough $H$, the magnetic gap of the side surface ensures a chiral hinge mode ($\frac{1}{2}+\frac{1}{2}$) at the top surface but nothing ($\frac{1}{2}-\frac{1}{2}$) at the bottom surface [see Figs. 4(a) and 4(b)]. In comparison, as $H$ reduces, for all side surfaces, the gap gradually becomes hybridization gap, leading to the localization of the half-quantized anomalous Hall conductivity at the top and bottom surfaces, as shown in Figs. 4(c) and 4(d).

For comparison, we also calculate the TME of the AFM AXI exemplified by A-type AFM MnBi$_2$Te$_4$ with the Néel vector oriented along the $z$-direction, manifesting a gapless Dirac-like side surface state because of the protection of $\mathcal{T}\tau_{1/2}$ [14, 15]. For a finite sample, the side surface gap is always a hybridization gap. Our calculations show that the TME coefficient is always zero as a function of the exchange coupling (Supplementary data Section 5). The comparison between the vanishing TME in the AFM AXI and the nonzero TME in the FM AXI reveals the importance of the surface magnetic gap in generating the topological response. We note that TME has been proposed in AFM magnetic heterostructure and MnBi$_2$Te$_4$ models under an in-plane electric field, which generates surface Hall currents circulating through the magnetic-gapped top and bottom surfaces and hybridization-gapped side surfaces [9, 10, 63]. However, it might be challenging to simultaneously maintain an appropriate surface gap size (requires a thin slab) and avoid interference between surface anomalous Hall currents from the top and bottom surfaces (requires a thick slab).

To summarize, we demonstrate that the inversion-preserved FM AXIs MnBi$_2$Te$_4$/(Bi$_2$Te$_3$)$_n$ serve as a realizable platform for achieving the topological magnetoelectric response. Designed in a triangular prism geometry, the side surfaces are gapped by a uniform FM exchange field instead of the elaborately oriented magnetic moments in previous studies. Using linear response theory, we obtain nearly half-quantized orbital magnetization induced by a vertical electric field. In a realistic finite-layer sample, the TME signal can be directly extracted from the itinerant circulation magnetization and local ion moments by reflecting the electric field. In parallel, an electric polarization induced by a time-dependent vertical magnetic field could also be expected in our configuration. Given the enormous number of observations verifying the successful topological band theory, our findings provide an accessible proposal to achieve the long-sought TME in realistic materials, which sheds light on more response properties predicted by topological field theory.

**Acknowledgements**

We thank Profs. Jianpeng Liu, Zhida Song and Chang Liu for helpful discussions. The numerical calculations were performed on high-performance computing systems supported by Center for Computational Science and Engineering of Southern University of Science and Technology.

**Funding**

This work was supported by National Key R&D Program of China under Grant No. 2020YFA0308900, Science, Technology and Innovation Commission of Shenzhen Municipality (No. ZDSYS20190902092905285), Guangdong Innovative and Entrepreneurial Research Team Program under Grant No. 2017ZT07C062, Guangdong Provincial Key Laboratory for Computational Science and Material Design under Grant No. 2019B030301001.

**Author contributions**

Q.L. conceived the idea and supervised the research. Y.W. and J.L. carried out the theoretical calculations and wrote the paper with Q.L.. Y.W. and J.L. contributed to the work equally.

**Conflict of interest statement**

None declared.

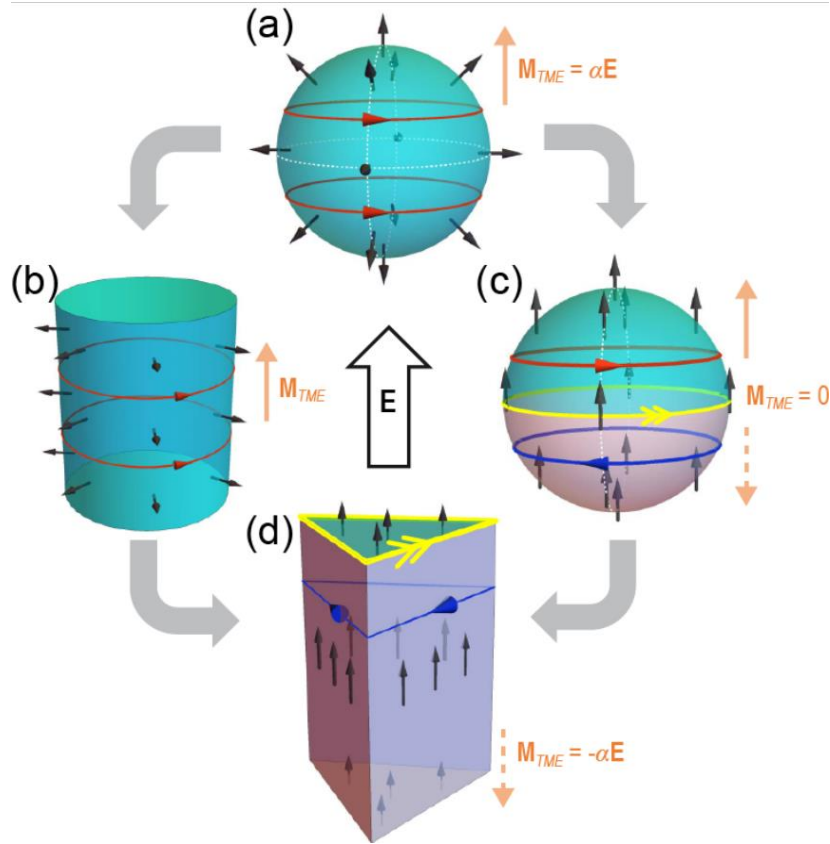

**Fig. 1.** (a) Sketch of a 3D spherical TI with hedgehog-like surface magnetization (black arrows). By applying an electric field **E**, the surface Hall effect gives rise to circulating current (red lines) and thus generates a bulk magnetization $\mathbf{M}_{TME} \parallel \mathbf{E}$, with $\alpha = \frac{e^2}{2hc}$ the fine structure constant. (b) The cylindrical TI with radial magnetic orbital moment, while the same bulk magnetization can be induced by **E**. (c) The spherical 3D TI with FM magnetization, where two hemispheres form two domains with a gapless chiral mode (yellow line). **E** cannot generate a net magnetization as the contributions from two domains compensate each other. (d) The triangular prism AXI such as FM MnBi$_2$Te$_4$, where the side surfaces are gapped due to the hexagonal warping effect. A net **E**-induced bulk magnetization is predicted.

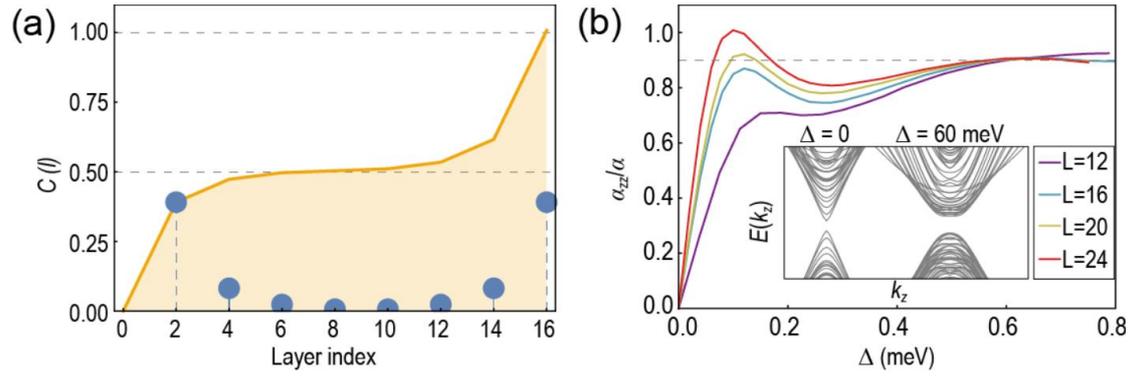

**Fig. 2.** (a) Layer-resolved Chern number of a 16-layer triangular prism FM MnBi$_2$Te$_4$. Blue points represent the local Chern number $\mathbb{C}(i)$ for each layer, and the orange line shows the integrated Chern number $C(l) = \sum_{i=1}^{l} \mathbb{C}(i)$ over the layers. (b) Magnetoelectric coefficient as a function of the exchange coupling $\Delta$ in triangular prism FM MnBi$_2$Te$_4$ with different side length $L$ and translation symmetry along the $z$-direction. Inset: band structures of the $L = 14$ configuration with different $\Delta$.

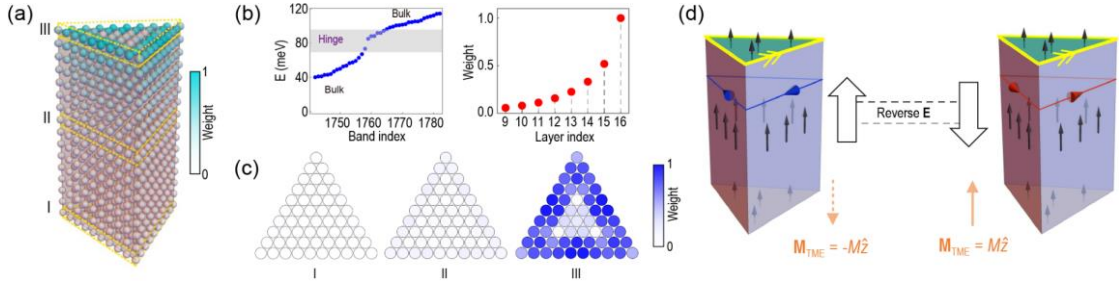

**Fig. 3.** (a) Lattice model of the triangle prism FM MnBi$_2$Te$_4$ for side length $L = 10$ and height $H = 16$. (b) Top panel: energy levels of the lattice model, with the eigenenergies of the hinge modes highlighted by gray shades. Bottom panel: layer distribution of one of the hinge modes. (c) In-plane distribution of the hinge mode for different layers in (a). (d) Switching the direction of the electric field reverses the Hall current and the sign of $\mathbf{M}_{TME}$, while the itinerant circulation magnetization originated from the chiral hinge mode remains intact.

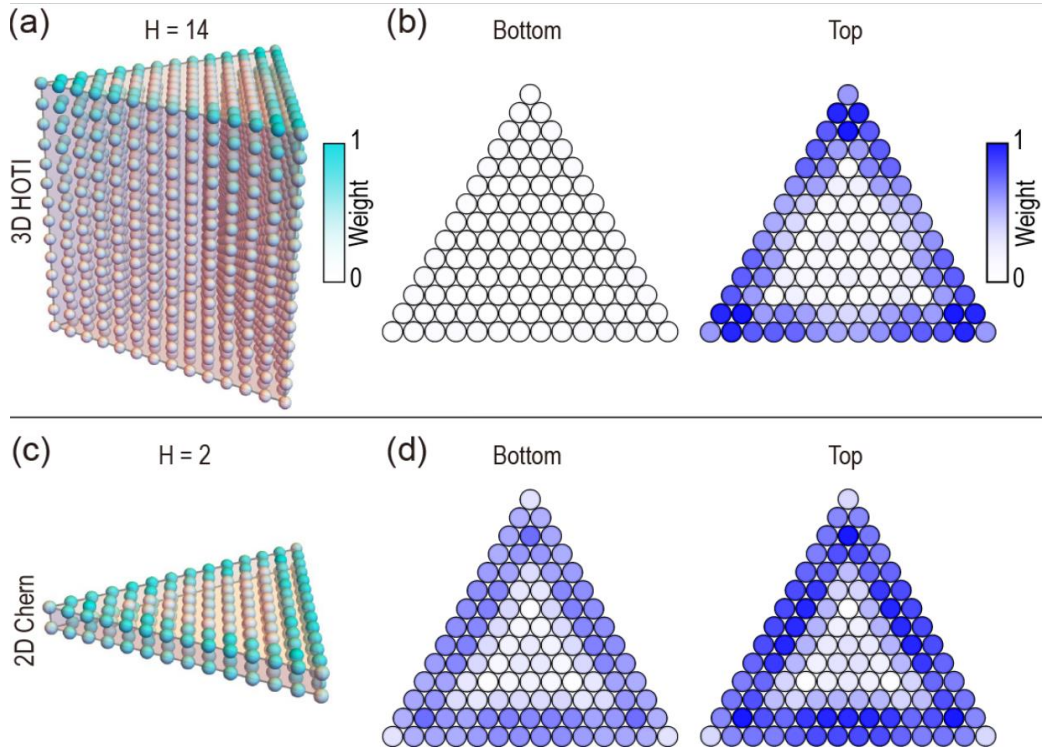

**Fig. 4.** (a) Lattice of a 14-layer triangle prism FM MnBi$_2$Te$_4$ as a 3D HOTI. (b) The distribution of the gapless hinge mode projected on the bottom and top layers of (a), respectively, indicating a well-localized hinge state. (c) Lattice of a bilayer structure of the same material, manifesting a 2D Chern insulator phase. (d) Projection of the gapless edge state on the bottom and top layers.